\documentstyle{l-aa}

\begin{document}

   \thesaurus{11.01.2 11.06.2 11.09.4 09.04.1 13.18.1}
   \title{Dust in high-$z$ radio-loud AGN\thanks
   {Partially based on observations obtained at the European Southern
   Observatory, La Silla, Chile}}

\subtitle{}

\author{A.\ Cimatti$^{1}$, W.\ Freudling$^{2,3}$, H.J.A.\ R\"ottgering$^{4}$, R.J.\ Ivison$^{5}$, P.\ Mazzei$^{6}$}
\institute{$^1$ Osservatorio Astrofisico di Arcetri, Largo E.\ Fermi 5,
		   I-50125, Firenze, Italy\\
		   $^2$ European Southern Observatory, Karl-Schwarzschild-Str.\ 2,
		   D-85748 Garching bei M\"unchen, Germany\\
		   $^3$ Space Telescope---European Coordinating Facility,
		   Karl-Schwarzschild-Str.\ 2,
		   D-85748 Garching bei M\"unchen, Germany\\
		   $^4$ Sterrewacht, Huygens Lab., P.O.\
		   Box 9513, 2300 RA Leiden, The
		   Netherlands\\
		   $^5$ Institute for Astronomy, University of Edinburgh,
                   Royal Observatory, Blackford Hill, Edinburgh EH9 3HJ, U.K.\\
		   $^6$ Osservatorio Astronomico,
		   Vicolo dell'Osservatorio 5, I-35122, Padova,
		   Italy}

   \offprints{A.\ Cimatti}

   \date{}

   \maketitle

   \begin{abstract}

We present continuum observations of a small sample of high-redshift,
radio-loud AGN (radio galaxies and quasars) aimed at the detection of
thermal emission from dust. Seven AGN were observed with IRAM and SEST
at 1.25\,mm; two of them, the radio galaxies 1243+036 ($z \sim 3.6$)
and MG\,1019+0535 ($z \sim 2.8$) were also observed at 0.8\,mm with
the JCMT submillimetre telescope.  Additional VLA observations were
obtained in order to derive the spectral shape of the synchrotron
radiation of MG\,1019+0535 at high radio frequencies. MG\,1019+0535
and TX\,0211$-$122 were expected to contain a large amount of dust based
on their depleted Ly$\alpha$ emission. The observations suggest a clear
1.25-mm flux density excess over the synchrotron radiation spectrum of
MG\,1019+0535, suggesting the presence of thermal emission from dust
in this radio galaxy, whereas the observations of TX\,0211$-$122 were
not sensitive enough to meaningfully constrain its dust content.  On
the other hand, our observations of 1243+036 provide a stringent upper
limit on the total dust mass of $<10^8$\,M$_{\odot}$.  Finally, we
find that the spectra of the radio-loud quasars in our sample ($2 < z
< 4.5$) steepen between rest-frame radio and the far-infrared. We
discuss the main implications of our results, concentrating on the
dusty radio galaxy, MG\,1019+0535.

      \keywords{
	  Galaxies: active --
	  Galaxies: fundamental parameters --
	  Galaxies: ISM --
	  ISM: dust --
	  Radio continuum: galaxies
               }
   \end{abstract}

\section{Introduction}

In recent years several attempts have been made to study the
interstellar medium in high-redshift galaxies. Promising results have
been obtained by millimetre and submillimetre observations, which
sample the rest-frame far-IR emission of galaxies at $z > 2$ (Andreani
et al.\ 1993).  The study of dust in distant galaxies is very
important because it provides information on the physical state of the
ISM, and its relation to other properties of the galaxies, such as
activity and evolution. For example, if it is correct to assume that
the rest-frame far-IR luminosity $L_{\rm FIR}$ is a measure of the
star-formation rate, then this could provide a way to estimate the
evolutionary state of high-$z$ galaxies. Establishing whether dusty
galaxies are common at high redshifts is also important in order to
evaluate the effects of dust obscuration on surveys of quasars and
protogalaxies carried out in optical bands (Smail, Ivison \& Blain
1997).

Several active galaxies with $z>2$ have been detected in
submillimetre/millimetre bands, suggesting the presence of large
amounts of dust in their host galaxies (M$_{\rm d} \sim 10^{8-9}$
M$_{\odot}$; see Hughes, Dunlop \& Rawlings 1997 for a recent review;
Omont et al.\ 1996b; Ivison et al.\ 1998). At high redshift, dust is
also thought to be present in damped Ly$\alpha$ absorption systems
(Pettini et al.\ 1994; Pei, Fall \& Bechtold 1991), in very red
galaxies (Hu \& Ridgway 1994), and in high-$z$ radio galaxies (Cimatti
1996 and references therein). A substantial amount of dust is also
expected in theoretical models of the evolution of galaxies at
high-$z$ (Mazzei \& De Zotti 1996 and references therein).  Finally,
it is important to recall that molecular gas has been observed in a
few distant active galaxies, allowing a direct estimate of the
dust-to-gas mass ratio at large cosmological distances for the first
time (see Omont et al. 1996a; Ohta et al.\ 1996).

We recall that particular caution is needed in the interpretation of
the submillimetre/millimetre data of radio-loud objects.  The
existence of a thermal dust emission excess should be confirmed by
checking whether the submillimetre/millimetre flux densities represent
a real excess over the synchrotron spectrum. The radio galaxy
B2\,0902+343 ($z\sim3.4$) is an example of the problem, where the
observed 1.3-mm flux density is not due to thermal emission from dust,
but to the tail of the radio synchrotron spectrum (Downes et al.\ 1996
and references therein). Therefore, high-frequency ($\sim8-43$\,GHz)
radio observations are extremely important to derive the shape of the
synchrotron spectrum and to estimate its contribution to the
millimetre region.

In the present paper we present the results of millimetric
observations of a small sample of radio galaxies and radio-loud
quasars with $z>2$, and additional high-frequency VLA observations of
MG\,1019+0535, a radio galaxy at $z\sim2.8$. Throughout this paper we
assume $H_0=50$\,km\,s$^{-1}$\,Mpc$^{-1}$, $q_0=0.5$ and define
$h_{50}=H_0/50$.

\section{Observations and data reduction}

\begin{table*}
\label{obs}
\caption{Summary of the observations.}
\begin{tabular}{lccclrcrccc} 
\hline
& & & & & & & & & & \\ 
Name & Class & $z$ & Tel & Date & $\lambda_{\rm obs}$ & $\lambda_{\rm rest}$ & t$_{\rm int}$ & $\tau$ & $S_{\nu}$ & $S_{\nu}/ \sigma$ \\ 
     &       &     &     &      & ($\mu$m)            & ($\mu$m)             & (s)           &        &  (mJy)    &  \\
  & & & & & & & & & & \\ \hline
  & & & & & & & & & & \\
TX 0211$-$122&RG&2.34&SEST&13 July 1995&1270&380&4500&0.19-0.22&-0.53$\pm$3.99&--\\
MRC 0943$-$242&RG&2.93&SEST&06 May 1996&1270&323&5500&0.11-0.13&4.33$\pm$3.29&1.3\\
MG 1019+0535&RG&2.76&IRAM&19-20 March 1996&1250&332&5760&0.30-0.40&2.13$\pm$0.47&4.5\\
MG 1019+0535&RG&2.76&JCMT&24 April 1996&790&210&10800&0.30-0.40&14.70$\pm$4.60&3.2\\
MRC 1043$-$291&RQ&2.13&SEST&07 May 1996&1270&406&800&0.15-0.17&29.38$\pm$8.15&3.6\\
1243+036&RG&3.58&JCMT&25 April 1996&790&172&14080&0.23-0.29&7.00$\pm$3.10&2.3\\
1243+036&RG&3.58&IRAM&19-20 March 1996&1250&273&2040&0.20-0.40&2.61$\pm$0.86&3.0\\
PKS 1251$-$407&RQ&4.46&SEST&07 May 1996&1270&233&3800&0.14-0.19&8.00$\pm$3.1&2.6\\
PKS 1354$-$107&RQ&3.00&SEST&06 May 1996&1270&317&2100&0.14-0.16&10.92$\pm$5.11&2.1\\
  & & & & & & & & & & \\ \hline
\end{tabular}
\footnotesize 

\vspace*{0.2cm}
Note: RG --- radio galaxy; RQ --- radio-loud quasar; t$_{\rm int}$ ---
integration time on the source; $\tau$ --- opacity during the observations.

\end{table*}

We have observed seven objects with $2<z<4.5$; four radio galaxies and
three radio-loud quasars. Table~1 shows the target details.  The radio
galaxies TX\,0211$-$122 and MG\,1019+0535 were observed because their
UV spectra are suggestive of extinction by dust (van Ojik et al.\
1994; Dey et al.\ 1995), MRC\,0943$-$242 because of the the presence
of a large amount of neutral hydrogen surrounding the host galaxy
(R\"ottgering et al.\ 1995) and 1243+036 because of its very high
redshift (van Ojik et al.\ 1996).  Among the quasars, we recall that
PKS\,1251$-$407 is, to date, the most distant known radio-loud quasar
(Shaver, Wall \& Kellermann 1996).  MRC\,1043$-$291 (Kapahi et al.\
1997) and PKS\,1354$-$107 (Shaver et al.\ 1997, in preparation), were
selected because of their convenient observability during the
observing run.

\subsection{SEST observations}

The observations with the 15-m SEST (Swedish---ESO Submillimetre
Telescope) were made at La Silla, Chile, during 1995 July and 1996
May. The telescope was equipped with a $^{3}$He-cooled 1.3-mm
bolometer, with a central frequency of 250\,GHz and a bandwidth of
around 50\,GHz.  The beamsize is about 24$''$ (FWHM) and the typical
sensitivity is 200\,mJy\,s$^{-1/2}$. The observations were carried out
during nights with rather low opacities. The opacity was checked by
several skydips taken during the nights (i.e.\ the telescope was moved
to six different elevations where the bolometer integrated for 10\,s
and a calculation of the zenith opacity was made).  The pointing was
checked using quasars as astrometric reference sources and the
absolute flux calibration was achieved by observing the planet Uranus.
We estimate that the flux calibration has a typical uncertainty of
20\%.  The observations were made in beam-switching mode with a
beam throw of 70$''$, and the data were reduced according to method
described by Andreani (1994) (see also Andreani et al.\ 1993). We have
found a linear decrease of the standard deviation with $t^{-1/2}$,
indicating that the observing conditions were stable and limited by
the sky noise.

\subsection{IRAM observations}

Observations with the IRAM 30-m antenna were made in 1996 March with
the MPIfR 1.3-mm 7-channel bolometer array. The bolometers are
separated by 22$''$ in an hexagonal arrangement surrounding the
central pixel. The individual beamsize is about 11$''$ (FWHM), and the
typical sensitivity of 60\,mJy\,s$^{-1/2}$. The ON--OFF technique uses
the wobbler with a beamthrow of 33$''$, integrating for 10\,s per
subscan. The opacity was monitored every 1---1.5\,hr with a skydip
procedure, and was around 0.4 during the first night and 0.2 during
the second. In both cases, the atmosphere was quite stable. The flux
calibration was achieved by observing Uranus, and the uncertainties
are again of order 20\%. The fluxes of the six outer channels were
averaged and subtracted from the central channel.

\subsection{JCMT observations}

The radio galaxies 1243+036 and MG\,1019+0535 were also observed at
the JCMT (the 15-m James Clerk Maxwell Telescope, Mauna Kea, Hawaii)
using the single-element $^{3}$He-cooled UKT14 bolometer (Duncan et
al.\ 1990), coupled with a broad-band filter ($\nu = 384$\,GHz;
$\Delta \nu = 30$\,GHz FWHM). The observations were made using a 65-mm
focal plane aperture, resulting in a FWHM beamwidth of 16.5$''$.
Sky emission was subtracted by chopping the secondary mirror in azimuth, 
at a frequency of 7.8\,Hz, with a throw of 60$''$.  

The radio galaxy 1243+036 was observed on 1996 April 26 during
excellent weather conditions. Calibration was taken from Mars (which
set near the beginning of the shift) and the secondary calibrators
NGC\,2071IR, CRL\,618, IRC~+10216 (CW Leo) and 16293$-$2422. 3C\,273
was used as a pointing source and bootstrap calibrator (observed eight
times). The 384-GHz zenith optical depth was typically around 0.29,
but it appeared to decrease towards the end of the shift (0.23). In
total, 880 pairs of 16\,s (i.e.\ 8\,s in each beam) were obtained. The
raw, uncalibrated dataset gave a S/N of 0.99. After despiking,
calibration and statistical testing, the final result was
$7.0\pm3.1$\,mJy. The calibrated 800-$\mu$m flux of 3C\,273 was
$8.2\pm0.1$\,Jy; IRC~+10216 (which is variable) was $5.4\pm0.1$\,Jy.

The radio galaxy MG\,1019+0535 was observed over a period of 5.5\,hr
during 1996 April 24. A total of 3\,hr was spent on-source, the
remaining 2.5\,hr being devoted to calibration. Flux calibration was
provided by regular measurements of IRC~+10216, as was a determination
of the 375-GHz zenith opacity. Observations of local pointing sources
kept rms pointing errors below 3$''$ and associated flux losses below
10\%; the overall uncertainty in the flux measurement was therefore
dominated by the poor S/N.  After careful editing, the MG\,1019+0535
measurements indicated a 375-GHz flux density of $14.7\pm4.6$\,mJy ---
a marginal detection.

\subsection{VLA observations}

MG\,1019+0535 was also observed at the Very Large Array (VLA), New
Mexico.  Radio continuum observations in the X (8.3\,GHz), U (15.0\,GHz) 
and K (23.0\,GHz) bands were carried out in the D configuration on 1996 
July 19 (with a bandwidth of 50\,MHz), and in the Q band using the A 
configuration on 1996 December 26. 

The total on-source time was 10 minutes each in the X and U bands, and
20\,m in the K band. The visibility averaging time was 10\,s in all
bands.  The phase was calibrated with calibrator source 1024$-$008
which was observed every 10\,m, and 3C\,286 was observed as a flux
calibrator.  The data were reduced using standard {\sc aips}
procedures.  The used synthesized beam sizes, are listed in Table~2.
Maps with a cell size of 0.5$''$ were produced for each band. Maps
were made and {\sc clean}ed using the {\sc mx} routine within {\sc
aips}.  The radio source was unresolved --- no evidence was found for
any extended emission. The total flux density, as measured from the
maps in each band, is also given in Table~2. The major source of
uncertainty for the measured flux density is the uncertainty in the
flux scale, which we estimate to be of order 10\%.

The Q band observation was carried out over a period of 1.7\,hr using
twelve VLA antennas. The total bandwidth for the observations was
100\,MHz, centred at 43.34\,GHz (6.9\,mm). Again, the observing and
calibration procedures were standard. After checking the pointing
accuracy of the antennas, brief observations of MG\,1019+0535 were
sandwiched and interspersed with measurements of 1055+018, a bright,
compact calibrator. The flux density of the galaxy was tied to that of
the calibrator (4.62\,Jy) which, in turn, was tied to the flux density
of 3C\,286 (1.49\,Jy on the shortest baselines).  The noise level at
the position of the galaxy agreed well with that measured using the
calibrated visibilities, giving a $3\sigma$ upper limit of 1.94\,mJy.

\begin{table}
\label{obs}
\caption{VLA observations of MG\,1019+0535.}
\begin{tabular}{lccccc} 
\hline
& & & & & \\ 
Date  & $\nu_{\rm obs}$ & $\nu_{\rm rest}$ & Flux & Config& Beam size FWHM\\ 
(1996)&    (GHz)        &   (GHz)          & (mJy)  & \& Band  & ($''$) \\
  & & & & & \\ \hline
  & & & & & \\
19 July&8.3&31.2&52.7$\pm$5.3&D,X&7.36$\times$5.97\\
19 July&15.0&56.4&23.2$\pm$2.3&D,U&4.31$\times$3.64\\
19 July&23.0&86.8&9.3$\pm$0.9&D,K&2.80$\times$2.52\\
26 Dec&43.3&162.8&3$\sigma<$1.94&A,Q&---\\
  & & & & & \\ \hline
\end{tabular}
\end{table}

\section{Results}

Table~1 shows the flux densities measured for our targets. The radio
galaxy MG\,1019+0535 is the only source detected at submillimetre
wavelengths at greater than the 4-$\sigma$ significance level.  In
this section we discuss the main implications of our observations for
each individual source. The dust masses are estimated using the
formula:

\begin{equation}
M_{\rm d} = {S(\nu_{\rm obs})D^{2}_{\rm L} \over
(1+z)\kappa_{\rm d}(\nu_{\rm rest})B(\nu_{\rm rest},T_{\rm d})} 
\end{equation}

\noindent
where $S$ is the flux density, $\nu_{\rm obs}$ and $\nu_{\rm rest}$
are, respectively, the observed and rest-frame frequencies, $D_{\rm
L}$ is the luminosity distance, $B$ is the black-body Planck function,
$T_{\rm d}$ is the dust temperature and $\kappa_{\rm d}= 0.67(\nu_{\rm
rest}/250\,{\rm GHz})^{2}$\,cm$^{2}$\,g$^{-1}$ is the adopted mass
absorption coefficient. In order to obtain limits on $M_{\rm d}$, it
is necessary to assume a temperature for the grains.  We assume a
representative temperature $T_{\rm d}$=60\,K, consistent with the
typical temperatures estimated for high-redshift active galaxies. For
a discussion about the uncertainties of dust masses and temperatures,
see Hughes et al.\ (1997). It should be emphasised here that, besides
the uncertainties on $T_{\rm d}$ and $\kappa_{\rm d}$, the absolute
dust masses are also strongly dependent on the choice of $H_0$ and
$q_0$; in fact, they change by a factor of 4 for values of $H_0$
ranging from 50 to 100\,km\,s$^{-1}$\,Mpc$^{-1}$, and by a factor of
around 2 if $q_0$ is changed from $q_0=0.1$ to $q_0=0.5$.

For each source we have searched NED\footnote{The NASA/IPAC
Extragalactic Database (NED) is operated by the Jet Propulsion
Laboratory, Caltech, under contract with the National Aeronautics and
Space Administration.} in order to derive the spectral energy
distributions (SEDs) over a broad range of frequencies.  If the error
of the flux density is unknown, we assume an uncertainty of 5\%.
Whenever the significance of the flux density is $\leq3\sigma$, we
provide an upper limit at the 3-$\sigma$ level. Whenever more than one
flux density value is available at the same frequency, we plot all the
available values. Figs.~1 and 2 and Table 3 show the SEDs derived for
our targets.

\begin{table*}
\label{obs}
\caption{Flux densities.}
\begin{tabular}{lccclccc} 
\hline
& & & & & & \\ 
Object & $\nu_{\rm obs}$ & Flux & Ref & Object & $\nu_{\rm obs}$ & Flux &
Ref \\ 
       &    (GHz)      & (mJy)  &    &              &    (GHz)      & (mJy)  & \\
  & & & & & & \\ \hline
  & & & & & & \\
\small
TX 0211$-$122&0.408&1020$\pm$40&L81&MRC 1043$-$291&0.408&1090$\pm$60&L81\\
			&1.465&189$\pm$9.5&C97&             &2.70&530$\pm$27&WO90\\
			&4.70&53.8$\pm$5.0&C97&            &4.85&714$\pm$51&W96\\
			&8.20&24.3$\pm$2.5&C97&            &5.00&320$\pm$16&WO90\\
			&230&$<$12&here&              &230&$<$24.3&here\\
			&4.61$\times10^{5}$&2.36$\pm$0.24$\times10^{-3}$&vO94&1243+036&0.178&3600$\pm$180&WO90\\
MRC 0943$-$242&0.408&1050$\pm$30&L81&  &0.408&1810$\pm$80&L81\\
            &1.50&245$\pm$13&C97&    &1.40&256$\pm$13.0&WB92\\
			&4.70&56$\pm$3&C97&      &2.70&120$\pm$6.0&WO90\\
			&4.85&44$\pm$11&Gri94&   &4.70&69.6$\pm$7.0&vO96\\
			&8.20&21$\pm$1&C97&      &8.30&28.2$\pm$3.0&vO96\\
			&230&$<$9.9&here&   &230&$<$2.6&here\\
			&4.61$\times10^{5}$&5.93$\pm$0.59$\times10^{-3}$&C97& &384&$<$9.3&here\\
MG 1019+0535&0.178&2100$\pm$105&WO90& &1.36$\times10^{5}$&1.28$\pm$0.13$\times10^{-2}$&vO96\\
			&0.365&925$\pm$31&D95& &4.61$\times10^{5}$&2.84$\pm$0.28$\times10^{-3}$&vO96\\
			&0.408&920$\pm$50&L81&PKS 1251$-$407&0.33&$<$400&S96\\
			&1.400&454$\pm$23&WB92& &1.40&260$\pm$10&S96\\
			&1.490&360$\pm$6&D95& &2.70&250$\pm$10&S96\\
			&2.700&180$\pm$9&WO90& &4.85&238$\pm$15&W94\\ 
			&4.850&100$\pm$12&Gri95& &4.85&237$\pm$16&Gre94\\
			&4.850&115$\pm$17&B91& &4.90&200$\pm$10&S96\\
			&4.850&132$\pm$19&Gre91& &5.00&220$\pm$10&WO90\\
			&8.333&52.7$\pm$5.3&here& &15.0&110$\pm$10&S96\\
			&8.439&59.1$\pm$1.5&D95& &8.40&150$\pm$10&S96\\
			&15.00&23.2$\pm$2.3&here& &230&$<$9.3&here\\
			&23.08&9.30$\pm$0.9&here& &3.75$\times10^{5}$&2.47$\pm$0.25$\times10^{-2}$&S96\\
			&43.30&$<$1.94&here& &6.82$\times10^{5}$&1.40$\pm$0.14$\times10^{-3}$&vO96\\
			&230.0&2.13$\pm$0.47&here&PKS 1354$-$107&2.70&260$\pm$13&WO90\\
                        &384.0&14.7$\pm$4.6&here& &4.86&284$\pm$14&Gri94\\
			&3000&$<$460&here& &5.00&270$\pm$14&WO90\\
			&5000&$<$200&here& &8.40&280$\pm$14&S97\\
			&1.20$\times10^{4}$&$<$170&here& &8.40&191$\pm$10&S97\\
			&2.50$\times10^{4}$&$<$110&here& &230&$<$15.3&here\\

			&1.36$\times10^{5}$&6.40$\pm$1.30$\times10^{-3}$&D95& & & \\
			&3.75$\times10^{5}$&1.70$\pm$0.17$\times10^{-3}$&D95& & & \\
			&4.61$\times10^{5}$&7.84$\pm$0.40$\times10^{-4}$&D95& & & \\
			&5.00$\times10^{5}$&5.05$\pm$0.50$\times10^{-4}$&D95& & & \\
			&6.00$\times10^{5}$&4.03$\pm$0.40$\times10^{-4}$&D95& & & \\\\
  & & & & & & \\ \hline
\end{tabular}
\footnotesize 

\vspace*{0.2cm}
Note: L81 --- Large et al.\ (1981); C97 --- Carilli et al.\ (1997);
vO94 --- van Ojik et al.\ (1994); Gri94 --- Griffith et al.\ (1994);
WO90 --- Wright \& Ostrupcek (1990); D95 --- Dey et al.\ (1995); WB92
--- White \& Becker (1992); Gri95 --- Griffith et al.\ (1995); B91 ---
Becker et al.\ (1991); Gre91 --- Gregory \& Condon (1991); W96 ---
Wright et al.\ (1996); vO96 --- van Ojik et al.\ (1996); S96 ---
Shaver et al.\ (1996); S97 --- Shaver et al.\ (in preparation); W94
--- Wright et al.\ (1994); Gre94 --- Gregory et al.\ (1994); Gri94 ---
Griffith et al.\ (1994).
\end{table*}

\subsection{Notes on individual sources}

\noindent
{\it TX\,0211$-$122}

The optical emission line ratios of this radio galaxy are highly
anomalous, with the flux of the Ly$\alpha$ line relatively weak, and
that of the N\,{\sc v}$\lambda$1240 line relatively strong compared to
those of C\,{\sc iv}$\lambda$1549, He\,{\sc ii}$\lambda$1640 and
C\,{\sc iii}]$\lambda$1909 (van Ojik et al.\ 1994).  This was
interpreted as being due to a strong starburst and a large amount of
dust. According to the 8.2-GHz flux density and spectral index
(R\"ottgering et al.\ 1994; Carilli et al.\ 1997), the expected
synchrotron flux density at 1.3\,mm is only $\sim$0.3\,mJy. Our SEST
observation yields a 3-$\sigma$ upper limit of $\sim$12\,mJy which
limits the total amount of dust, according to equation (1), to $<1.1
\times 10^{9}$\,M$_{\odot}$. Although this limit is not very
stringent, it does suggest that the total amount of dust cannot be
much larger than that of the most dusty known active galaxies: for
instance, the quasar BRI\,1202$-$0725 has $M_{\rm d} \sim
10^{9}$\,M$_{\odot}$ (Isaak et al.\ 1994; Hughes et al.\
1997). Finally, we recall that the amount of molecular hydrogen
estimated from observations of CO is $<10^{11}$\,M$_{\odot}$; not much
greater than that of nearby gas-rich starburst galaxies (van Ojik et
al.\ 1997).

\noindent
{\it MRC\,0943$-$242}

The detection of a halo of neutral hydrogen linked with the host
galaxy of MRC\,0943$-$242 and the dust which might be associated with
the neutral ISM (R\"ottgering et al.\ 1995) prompted the SEST
observation of this source. We recall that the limit on the molecular
hydrogen mass is $<10^{11}$\,M$_{\odot}$ (van Ojik 1995). According to
the 8.2-GHz flux density and spectral index (R\"ottgering et al.\
1994; Carilli et al.\ 1997), the expected synchrotron flux density at
1.3\,mm is only $\sim$0.2\,mJy. Our 1.3-mm 3-$\sigma$ upper limit of
9.9\,mJy limits the mass of dust to $<6.8 \times
10^{8}$\,M$_{\odot}$).  Because of the large uncertainty in the total
gas content (H+H$_2$) (R\"ottgering et al.\ 1995; van Ojik et al.\
1997), it is not possible to meaningfully constrain the dust/gas mass
ratio in this galaxy.

\noindent
{\it MG\,1019+0535}

This radio galaxy has spectroscopic properties similar to
TX\,0211$-$122, again indicating the possible presence of dust (Dey et
al.\ 1995). Our IRAM observations provided a suggestive detection at
1.25\,mm, and our JCMT observations yielded a marginal detection at
800\,$\mu$m. The IRAM observations were split into two nights but in
this case, unlike 1243+036, MG\,1019+0535 gave consistently positive
signal. Although the result is formally significant, we consider that 
our 1.25-mm data provide only a tentative detection because of the 
very weak flux density. We note, however, that previous IRAM detections 
at around this level have since proved to be trustworthy --- that of 
8C\,1435+635, for example (Ivison 1995; Ivison et al. 1998). 

It is important to stress that there are major
uncertainties in the interpretation of the millimetric observations of
this galaxy.  Optical imaging shows the presence of two objects
separated by about 1.5$''$ --- object A, identified as the counterpart
of the radio source at $z=2.76$, and object B (Dey et al.\ 1995). The
nature of B is unclear: it may be physically related to A, or be a
foreground galaxy at $z\sim0.66$. The problem is that the beam widths
of our 1.25-mm and 800-$\mu$m observations include both
objects. However, if the two objects are unrelated, the depression of
the Ly$\alpha$ line favours component A being the dusty object and the
source of the observed flux density at 1.3\,mm.

In order to better constrain the SED of this galaxy, we used data from
{\em IRAS}. However, since MG\,1019+0535 is not detected, upper limits
have been estimated at 12, 25, 60 and 100\,$\mu$m by searching a 1
square degree field centred on MG\,1019+0535 for sources from the {\em
IRAS Faint Source Catalogue}, adopting the faintest in each band as
the upper limit (0.11, 0.17, 0.20 and 0.47\,Jy, respectively). This
crude method relies on the fact that if the {\em FSC}'s sophisticated
search routines cannot find a point source, then the source must be
below the 3-$\sigma$ threshold. The method is less prone than some to
providing misleadingly low limits (Ivison 1995). The implication of
this result is dicussed in detail in Section~4.

\noindent
{\it MRC\,1043$-$291}

This radio-loud quasar has radio flux densities of 1.09 and 0.68\,Jy
at 408\,MHz and 5\,GHz, respectively (Kapahi et al.\ 1997). Therefore,
if we adopt a spectral index $\alpha=-0.19$ (defined as $S_{\nu}
\propto \nu^{\alpha}$), we derive an expected 1.3-mm synchrotron flux
density of around 326\,mJy. Our SEST observation provides a flux
density around an order of magnitude lower than expected, suggesting
that the radio spectrum steepens rapidly at high frequencies.

\noindent
{\it 1243+036 (= 4C\,03.24)}

This is the radio galaxy with the highest redshift in our observed
sample.  One spectacular feature is the presence of a Ly$\alpha$ halo
(with a luminosity of $\simeq 10{^{44}}$\,erg\,s$^{-1}$) which extends
over 20$''$ ($\simeq 135$\,kpc) (van Ojik et al.\ 1996).  The
Ly$\alpha$ images, coupled with high-resolution spectra, indicate that
the radio jet is interacting vigorously with the gas in the inner
region.  Perhaps most surprising is the low-surface-brightness outer
region of the Ly$\alpha$ halo.  Deep spectroscopy shows that it is
relatively quiescent ($\simeq 250$\,km\,s$^{-1}$ FWHM), but that there
is a velocity gradient of 450\,km\,s$^{-1}$ over the extent of the
emission ($\simeq 135$\,kpc). Because the halo extends beyond the
radio source, it is probable that its kinematics must predate the
radio source. The ordered motion may be a large-scale rotation caused
by the accretion of gas from the environment of the radio galaxy or by
a merger.

The extrapolation of the radio flux density at 8.3\,GHz (R\"ottgering
et al.\ 1994; van Ojik et al.\ 1996) provides expected synchrotron
flux densities of 0.2 and 0.4\,mJy, respectively, at 800\,$\mu$m and
1.3\,mm. We observed this source, both with the IRAM telescope and the
JCMT, but we obtained only non-significant detections at the
2---3-$\sigma$ level (see Table~1). The IRAM observations were
performed on two different nights. Although the combined observations
of the two nights provide a formal 3-$\sigma$ detection, we found that
this result is not reliable because the source was detected only
during the first night. In fact, deeper observations with the IRAM
telescope failed to detect the galaxy and provided a 3-$\sigma$ upper
limit $<1.5$\,mJy (R.\ Chini, private communication).

However, we can see that our 3-$\sigma$ upper limits ($<$9.3\,mJy at
800\,$\mu$m; $<$2.6\,mJy at 1.3\,mm) provide a relevant result. In
fact, the inferred total dust masses are $<1.0 \times 10^{8}$ and
$<1.3 \times 10^{8}$\,M$_{\odot}$ using the 800-$\mu$m and 1.3-mm
upper limits, respectively. The most stringent limit on the dust mass
(provided by the JCMT observation) implies that the amount of dust in
1243+036 is lower than that inferred for those high-$z$ radio
galaxies and quasars so far detected (see Hughes et al.\ 1997 and
references therein). The upper limit on $M_{\rm d}$ can be lowered
still further if we use the limit provided by Chini at 1.3\,mm, which
gives $M_{\rm d} < 7.7 \times 10^{7}$\,M$_{\odot}$. It is also
important to notice that CO observations of this galaxy have provided
a stringent limit on the amount of molecular hydrogen $<5 \times
10^{10}$\,M$_{\odot}$ (van Ojik 1995).  

\noindent
{\it PKS\,1251$-$407}

To date, this is the furthest known radio-loud quasar (Shaver et al.\
1996).  Our SEST observation provides a tentative (2.6-$\sigma$)
detection. Figure~2 shows the SED of this quasar. The 1.3-mm upper
limit hints that the synchrotron spectrum steepens at high
frequencies, as do the other two radio-loud quasars, MRC\,1043$-$291
and PKS\,1354$-$107.

\noindent
{\it PKS\,1354$-$107}

The optical counterpart of this radio source was identified by Shaver
et al.\ (P.A.\ Shaver, J.V.\ Wall, K.I.\ Kellermann, C.A.\ Jackson,
M.R.S.\ Hawkins, private communication) with a quasar at $z=3.0$. The
available radio flux densities (see Figure~2) suggest the presence of
a very flat spectrum from 2.7 to 8.4\,GHz. Our SEST observation
provides a 3-$\sigma$ upper limit of $<$15.3\,mJy, implying a sharp
steepening of the synchrotron spectrum at high frequencies.

\section{The case of MG\,1019+0535: a dusty radio galaxy}

Although the JCMT observation provides only a marginal detection, 
its combination with the IRAM detection of an excess over the 
synchrotron spectrum strongly suggests the presence of thermal dust 
emission from this galaxy (Fig.~3).

Hughes et al.\ (1997) have demonstrated that the uncertainties in
calculating the mass of dust responsible for the optically thin,
thermal, submillimetre emission are: (i) our limited knowledge of the
rest-frame mass absorption coefficient, $\kappa_{\rm d}$, and how this
quantity varies with frequency; (ii) the dust temperature ($T_{\rm
d}$), and, finally, (iii) the unknown values of $H_0$ and
$q_0$. Unfortunately, these problems are coupled. For example, Hughes
et al.\ (1993) noted that there is a trade off between $T_{\rm d}$ and
the critical frequency at which the dust becomes optically thick,
$\nu_0$.

Typically, each of these uncertainties can account for changes of up
to a factor $\sim 5$ in derived dust masses (see Hughes et al.\ 1997
for more details).  However, a conservative value for the dust mass
can be obtained if the paramaters we use to estimate the dust mass are
taken such that the dust mass is minimised. We further note that
(since for high-$z$ objects the submillimetre observations sample the
Rayleigh-Jeans region) the slope of the dust spectrum is not a
function of temperature.

Our measurements of MG\,1019+0535 at 240 and 384\,GHz suggest that the
submillimetre spectral index ($\alpha$, where $S_{\nu} \propto
\nu^{\alpha}$) is large and positive ($\alpha = +4.2 \pm 1.2$).  We
recall that the maximum allowed spectral index for self-absorption is
+2.5. Our result therefore rules out the possibility that the emission
is due to self-absorbed synchrotron radiation (Chini et al.\
1989). However, given the uncertainty of the 384-GHz flux density,
data at more frequencies are needed to better constrain $\alpha$.

Strong support for the thermal nature of the submillimetre emission
is provided by our deep measurements at 22 and 43\,GHz using the
VLA. These show that the steepening centimetre spectral index (Figs.~1
and 3) becomes still more negative as it approaches the millimetre
domaine; the predicted contribution at 240\,GHz from the dominant
centimetre component lies several orders of magnitude below the
measured 240-GHz flux density.

At first sight this indicates that the frequency dependence of the
dust grain emissivity (or the emissivity index, $\beta$) is $+2.2 \pm
1.2$, which encompasses the range normally quoted for interstellar
grains ($1.0 < \beta < 2.0$) as well as some less physical values
($\beta > 2.0$). However, the redshift is high and the rest-frame
frequency of the observed 374-GHz emission is close to the turnover of
the dust spectrum, so we do in fact require a high value of $\beta$ to
fit both the 240- and 384-GHz data. For $\beta=+2.0$, we find that $35
< T_{\rm d}/{\rm K} < 180$. The lowest temperatures (35\,K) are found
for an optically thin solution; the highest temperature (180\,K) is
permitted when we allow the dust to become optically thick (say at
$\nu_0 = 1.5$\,THz or 200\,$\mu$m) and to be constrained by the {\em
IRAS} upper limits.

Although it is clear that our observations do not constrain
stringently the dust temperature, we can make use of the usual theory
(Hughes et al.\ 1997) to estimate the mass of dust responsible for the
emission detected by IRAM and JCMT. For $35 < T_{\rm d}/{\rm K} <
180$, $0.17 < M_{\rm d}/10^8\,{\rm M}_{\odot} < 1.8$.

It is reassuring that $T_{\rm d} = 40$\,K is viable since this is the
temperature of the dust measured in the $z=4.25$ radio galaxy,
8C\,1435+635 (Ivison et al.\ 1998).  For $T_{\rm d} = 40$\,K, adopting
the same dust parameters as Ivison et al., we derive $M_{\rm d} = 2
\times 10^8 h_{50}^{-2}$\,M$_{\odot}$ which, when compared with the
dust mass estimate of $4 \times 10^8 h_{50}^{-2}$\,M$_{\odot}$ for
8C\,1435+635, suggests that powerful radio galaxies evolve
significantly between $z=4.25$ and $2.76$ (though we note that the
rest-frame 6-cm luminosity of 8C\,1435+635 is around 5 times that of
MG\,1019+0535 and that observations of complete samples of radio
galaxies spanning a range of redshifts and radio luminosities will be
required to trace their evolution in detail).

\subsection{Modelling the UV--to--FIR SED}  

The interpretation of the observed spectral energy distributions
(SEDs), is not straightforward, since a non-thermal contribution
cannot be neglected and the commonly used population synthesis models
do not allow for dust extinction. Using the same approach followed by
Mazzei \& De Zotti (1996), based on chemo-photometric population
synthesis models incorporating extinction and re-emission by dust and
accounting for non-thermal emission, we have attempted to analyse the
SED of MG\,1019+0535. We recall here that Dey et al.\ (1995) estimate
an upper limit to the internal reddening, $E_{B-V}<0.43$\,mag.

Our IRAM (and, marginally, JCMT) observations strongly favour the
presence of dust.  Although our spectral coverage is rather poor, we
attempt to fit the overall SED of this galaxy, from the optical to
1.25\,mm in the observed frame, with the aim of constraining the
evolutionary properties of MG\,1019+0535. For this study, we assume
that the source of the submillimetre continuum radiation is component
A (as suggested by its depleted Ly$\alpha$ emission), and we adopt its
optical and near-IR fluxes accordingly (Dey et al.\ 1995).

We have computed several models with Salpeter's initial mass function
(IMF) and different lower mass limits, $m_l$, as described in Mazzei
\& De Zotti (1996) (and references therein).  For a given model we
derive the age of the system which matches the data, with different
amounts of non-thermal AGN emission. We find an interesting result:
these data are well matched by models which always correspond to a
0.8---1-Gyr-old host galaxy, accounting for a non-thermal contribution
ranging from 50 to 90\% of the total flux density at 0.6\,$\mu$m (see
Fig.~4). According to this result, MG\,1019+0535 cannot be considered
a ``primaeval'' galaxy candidate because the bulk of its stellar
population is significantly evolved.  For
$H_0=50$\,km\,s$^{-1}$\,Mpc$^{-1}$, the formation redshift, $z_{\rm
form}$, of MG\,1019+0535 is between 10 and 4 if $q_0=0.5$ and $<5$ if
$q_0=0$.  In the following we will refer to the first value of $q_0$.

The expected bolometric luminosity is always larger than
$10^{13}\,L_{\odot}$; in particular, this rises by a factor 2.5 if
$m_l=0.5\,M_{\odot}$.  This corresponds to a residual gas fraction of
2\%, i.e.\ to a total gas mass of about $4\times10^{10}$\,M$_{\odot}$,
which is well below the Evans et al.\ (1996) upper limit, with
$E_{B-V}=0.13$, a total barionic mass, $M_{\rm barion}$, of around
$1.7\times10^{12}$\,M$_{\odot}$, and a star-formation rate of
800\,M$_{\odot}$\,yr$^{-1}$.  In this scenario, the hot stars are
almost completely obscured by dust which, heated by their radiation
field, transfers their bolometric luminosity to the far-IR wavelength
regime. Models with lower $m_l$ require larger $M_{\rm barion}$ and
higher star-formation rates. We derive $E_{B-V} \le 0.26$ for a
residual gas fraction as large as 30\% --- the largest allowed by
models --- and $M_{\rm barion} < 4.5 \times 10^{12}\,M_{\odot}$.

The available data can be fully accounted for by opaque models like
those already used by Mazzei \& De Zotti (1994) to fit the spectrum of
the ultraluminous galaxy IRAS\,$F10214+4724$. However, there is still
considerable latitude for modelling. Crucial constraints may be
provided by ground-based submillimetre measurements and by
observations with the {\em Infrared Space Observatory (ISO)}; these
measurements will help to define the shape of the far-IR SED, so
settling the dust temperature, and the role of PAHs in the near-IR
spectral range.

\section{Results on radio-loud quasars}

The three quasars show a sharp steepening of their synchrotron spectra
between rest-frame radio and submillimetre wavelengths. It is
important to recall that variability may affect the flux densities of
radio-loud 
quasars obtained at different epochs. In order to estimate the
magnitude of the spectral break in the millimetre region, we evaluate
the expected synchrotron flux density at $\nu_{\rm obs}$=230\,GHz,
deriving the spectral index from the available data and extrapolating
the 5-GHz radio flux densities to 1.3\,mm. If two flux densities are
available at the same frequency, we adopt the average value. For
PKS\,1251$-$407, we use only the quasi-simultaneous 1.4---15-GHz data
of Shaver et al.\ (1996).  For MRC\,1043$-$291, we find that
$\alpha(0.408-5\,{\rm GHz})=-0.32$, and that the expected flux density
at 230\,GHz is $S_{\rm exp}(230)\sim152$\,mJy. For PKS\,1251$-$407 and
PKS\,1354$-$107, we find $\alpha(1.4-15\,{\rm GHz})=-0.37$,
$\alpha(2.7-8.4\,{\rm GHz})=-0.07$, with $S_{\rm exp}(230)\sim50$\,mJy
and $S_{\rm exp}(230)\sim210$\,mJy, respectively. If we compare the
expected flux densities with the observed upper limits, we find that
$S_{\rm exp}(230)/S_{\rm obs}(230)>5-14$, i.e.\ the observed flux
density can be more than an order of magnitude lower than expected. We
cannot know what fraction of this cut-off {\it may} be due to
variability; however, the typical maximum amplitude of the variability
of flat-spectrum radio sources at $\lambda\sim$ 1\,mm---11\,cm, over
timescales of months---years, is 40---60\% (Jones et al.\ 1981
and references therein). It therefore seems unlikely that variability
alone can explain the sharp cut-offs observed in the three quasars in
our sample.

Moreover, high-frequency spectral turnovers have been often observed
in radio-loud quasars at lower redshifts (Chini et al.\ 1989;
Antonucci, Barvainis \& Alloin 1990; Klein et al.\ 1996). In
particular, the spectral break occurs in radio-loud quasars with
either flat or steep radio spectra, whereas in BL~Lac objects the
submillimetre flux densities lie on an extrapolation of the radio
spectrum (Knapp \& Patten 1991). Our observations show a turnover at
log$\nu_{\rm rest} ({\rm Hz}) \leq 11.8-12$, but the lack of further
spectral information does not allow us to infer whether the steepening
of the synchrotron--sub-mm spectra
is related to thermal emission from dust peaking at the higher
frequencies. It is important to recall that the question of dust in
high-redshift, radio-loud quasars is very relevant to unification
models (see Baker \& Hunstead 1995 for a discussion of dust in
radio-loud quasars) and to test the hypothesis that many quasars may
be missed in optical surveys (see Webster et al.\ 1995). Future
observations with {\em ISO} and SCUBA will extend the SED coverage to
higher frequencies for a large number of objects.

\section{Concluding remarks}

We have presented observations of the rest-frame far-infrared
continuum of a small sample of radio-loud AGN (4 radio galaxies and 3
quasars) with $2<z<4.5$.

One of the main findings is the detection of thermal continuum
emission from the radio galaxy MG\,1019+0535 ($z\sim2.8$). This is
particularly important since it confirms the suggestion that its weak
Ly$\alpha$ emission is probably due to dust extinction. We attempt to
estimate the temperature of the dust in order to derive the dust total
mass, but the present data data do not allow us to constrain
stringently the range of possible temperatures.  However, we estimate
that the total dust mass should be of order $0.2-2.0 \times
10^{8}$\,M$_{\odot}$ for temperatures $T_{\rm d} \sim 35-180$\,K. The
overall rest-frame UV--FIR SED can be accounted for by a relatively
evolved host galaxy (age $\sim$0.8\,Gyr) experiencing an episode of
vigorous star formation.

The radio galaxy 1243+036 ($z\sim3.6$), on the other hand, seems to
have a lower amount of dust and molecular gas than those active
galaxies that have so far been detected at high redshifts, indicating
that the properties of the ISM in active objects at $z>2$ can be
rather inhomogeneous.

Finally, the three flat-spectrum radio-loud quasars ($2<z<4.5$)
observed at 1.3\,mm show evidence of a spectral turnover at high
frequencies. Our data do not allow us to understand what fraction of
the turnover is due to the effects of variability, but we conclude
that variability cannot be the only cause.

\begin{figure*}
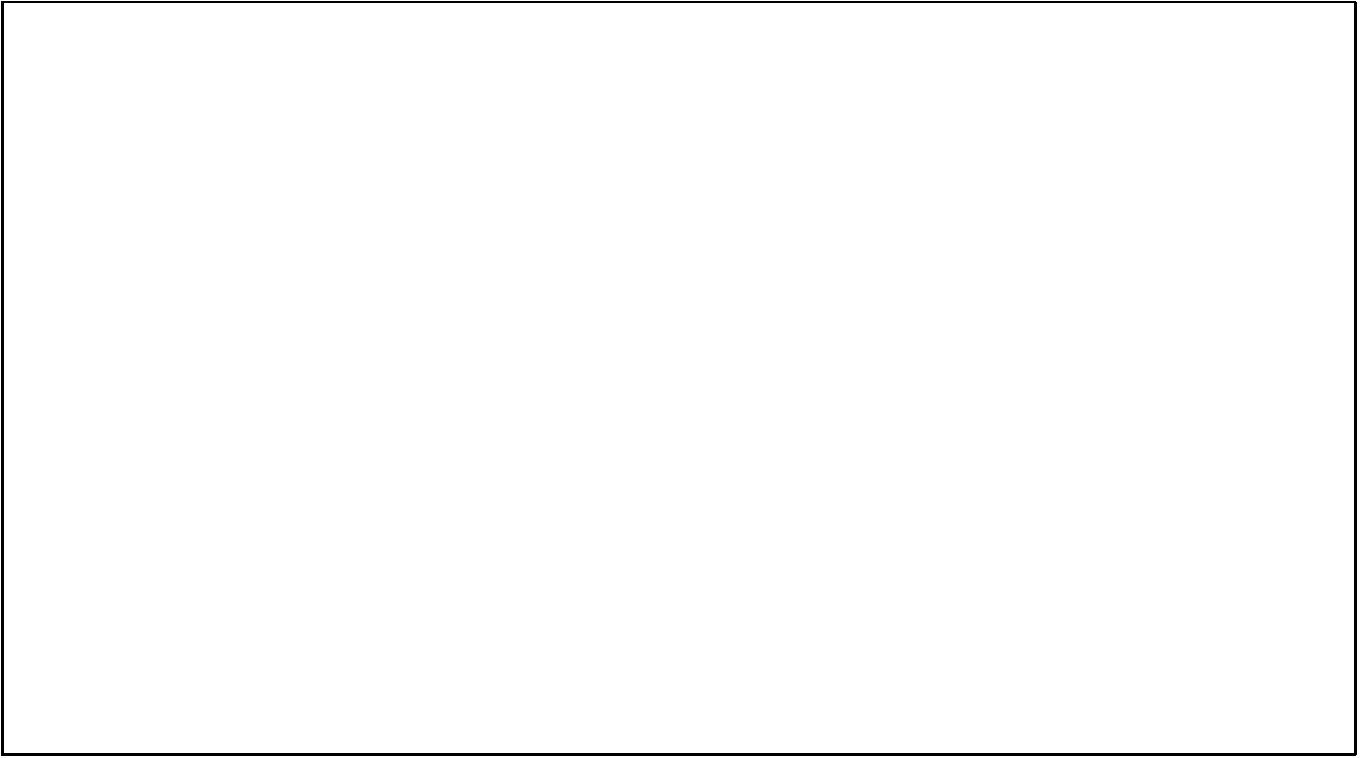

\picplace{10cm}
\caption{Spectral energy distributions of radio galaxies. The flux densities
and the relative references are listed in Table 3.
} 
\end{figure*}

\begin{figure*}
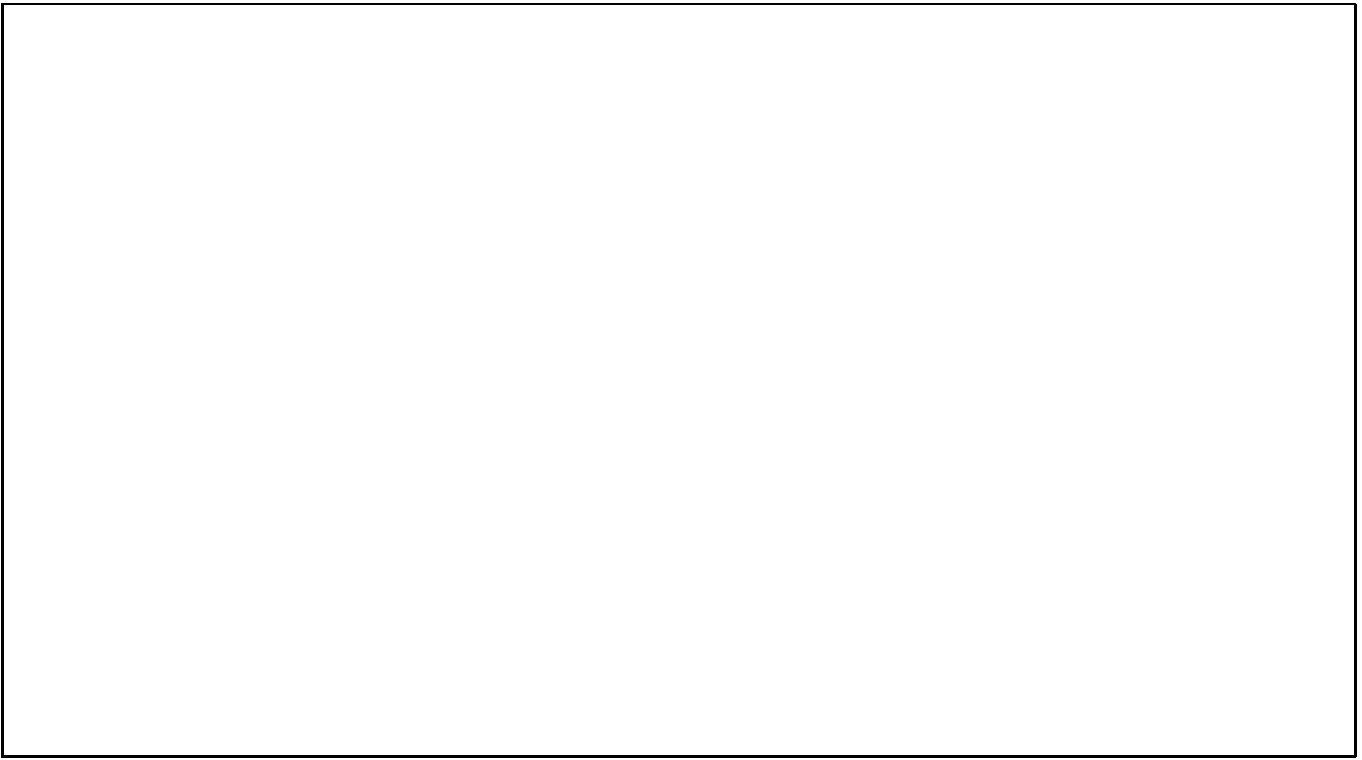

\picplace{10cm}
\caption{Spectral energy distributions of radio-loud quasars.
The flux densities and the relative references are listed in Table 3.
} 
\end{figure*}

\begin{figure*}
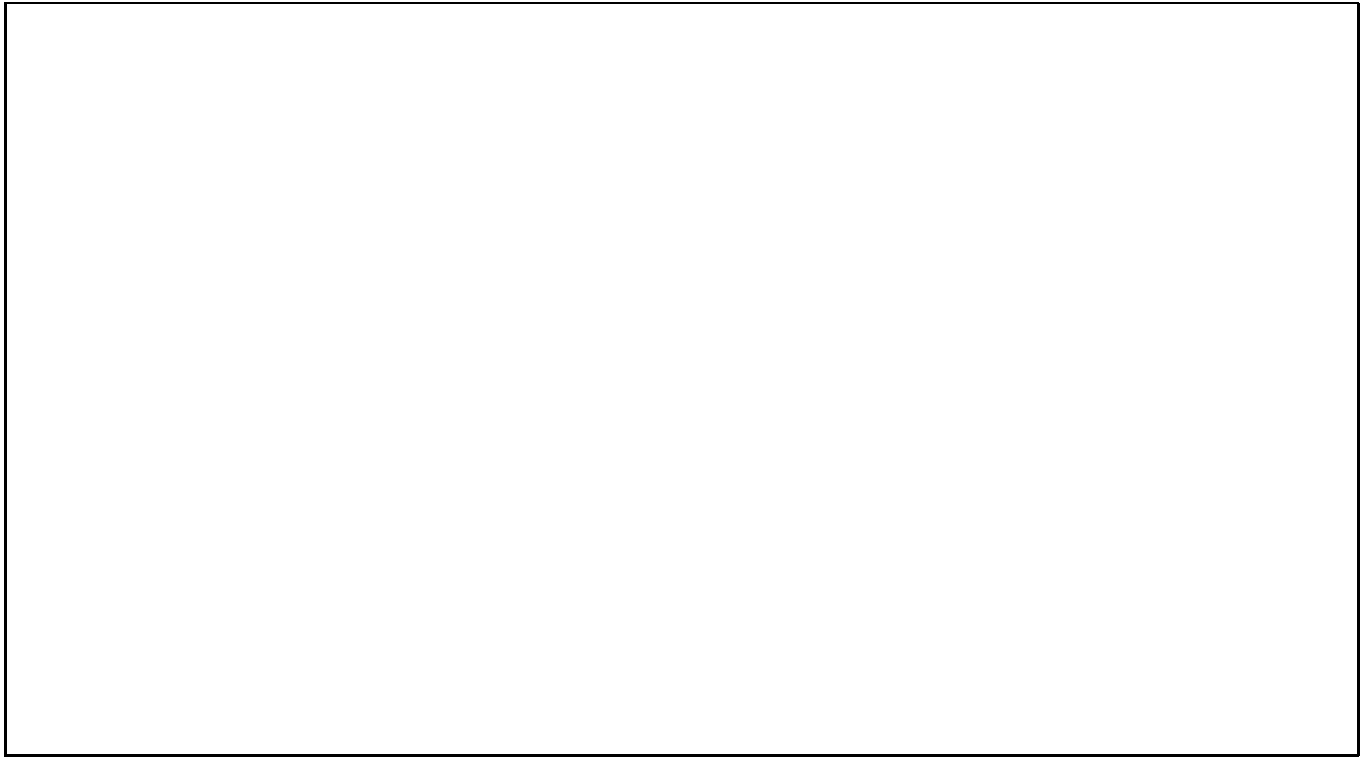

\picplace{10cm}
\caption{Spectral energy distribution of MG\,1019+0535. The lines
drawn in the rest-frame far-IR are modified 35- and 180-K blackbodies,
with a $\beta$=+2 frequency dependence for the dust-grain emissivity,
representing the most extreme dust temperatures compatible with our
data. The 180-K blackbody is optically thick at 200\,$\mu$m. Key:
circles --- VLA, JCMT, IRAM and {\em IRAS} measurements described in
the text; diamonds --- measurements from Dey et al.\ (1995); squares
--- measurements from Griffith et al.\ (1995), Becker, White \&
Edwards (1991), Wright \& Otrupcek (1990) and White \& Becker (1992).}
\end{figure*}

\begin{figure*}
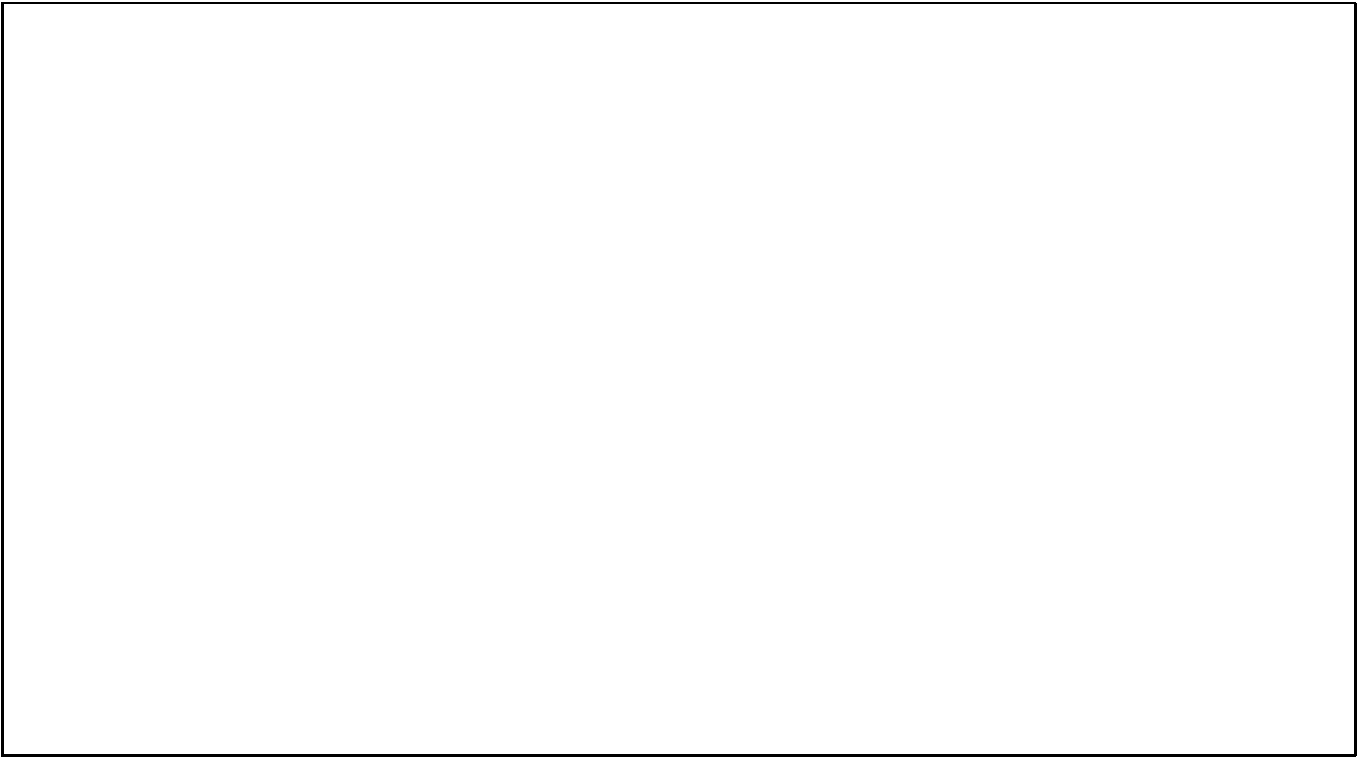

\picplace{10cm}
\caption{The fits to the SED of MG\,1019+0535 as obtained with the Mazzei \& De
Zotti (1996) model. Thick lines correspond to the overall match of the
SED of MG\,1019+0535 for models including a non thermal contribution,
stellar emission and dust effects (see text); a) shows the results for
two models corresponding to a host galaxy 1-Gyr old, short-dashed
($m_l=0.01\,M_{\odot}$) and long-dashed lines ($m_l=0.5\,M_{\odot}$),
with a non-thermal contribution at $0.64\,\mu$m of $70$ and $80\%$
respectively (thin line); the overall match for a host galaxy 0.8-Gyr
old (dot-dashed line, $m_l=0.1\,M_{\odot}$) with a non-thermal
contribution of $50\%$ at the same wavelength is also shown; in b) are
the results for the same models raising the non-thermal contribution at
the same wavelength to $90\%$; short-dashed curves require a dust
temperature of 46\,K instead of 60\,K.  }
\end{figure*}

\begin{acknowledgements}

We acknowledge the referee, A. Omont, for the useful suggestions.
We thank Raphael Moreno and Goeran Sandell who did some of the
observations in service mode at the IRAM and JCMT telescopes
respectively, and Roberto Fanti for useful discussions. This work was
supported in part by the Formation and Evolution of Galaxies network
set up by the European Commission under contract ERB FMRX-CT96-086 of
its TMR programme and by the high-$z$ programme subsidy granted by the
Netherlands Organization for Scientific Research (NWO). RJI is
supported by a PPARC Advanced Fellowship.

\end{acknowledgements}

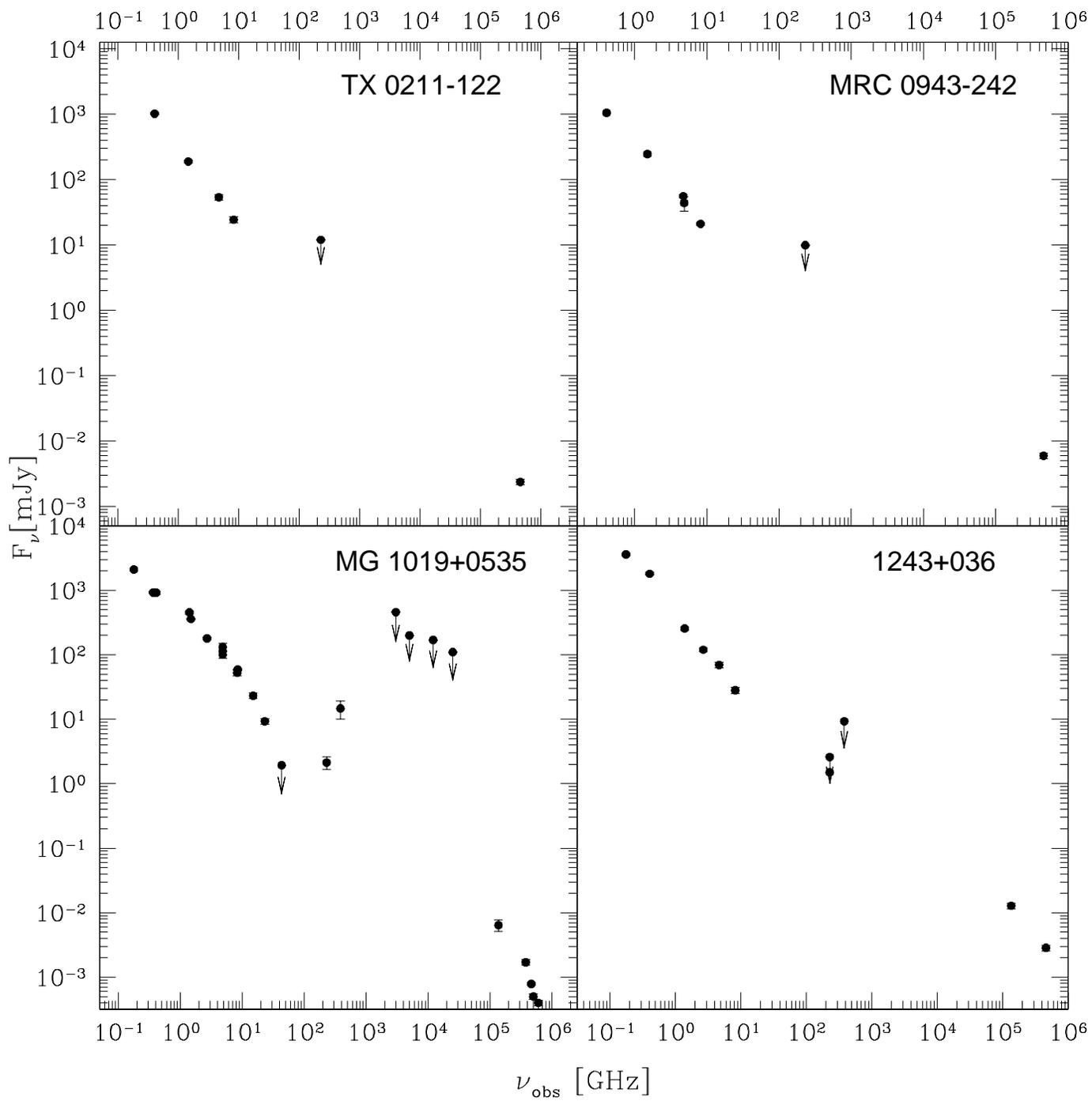
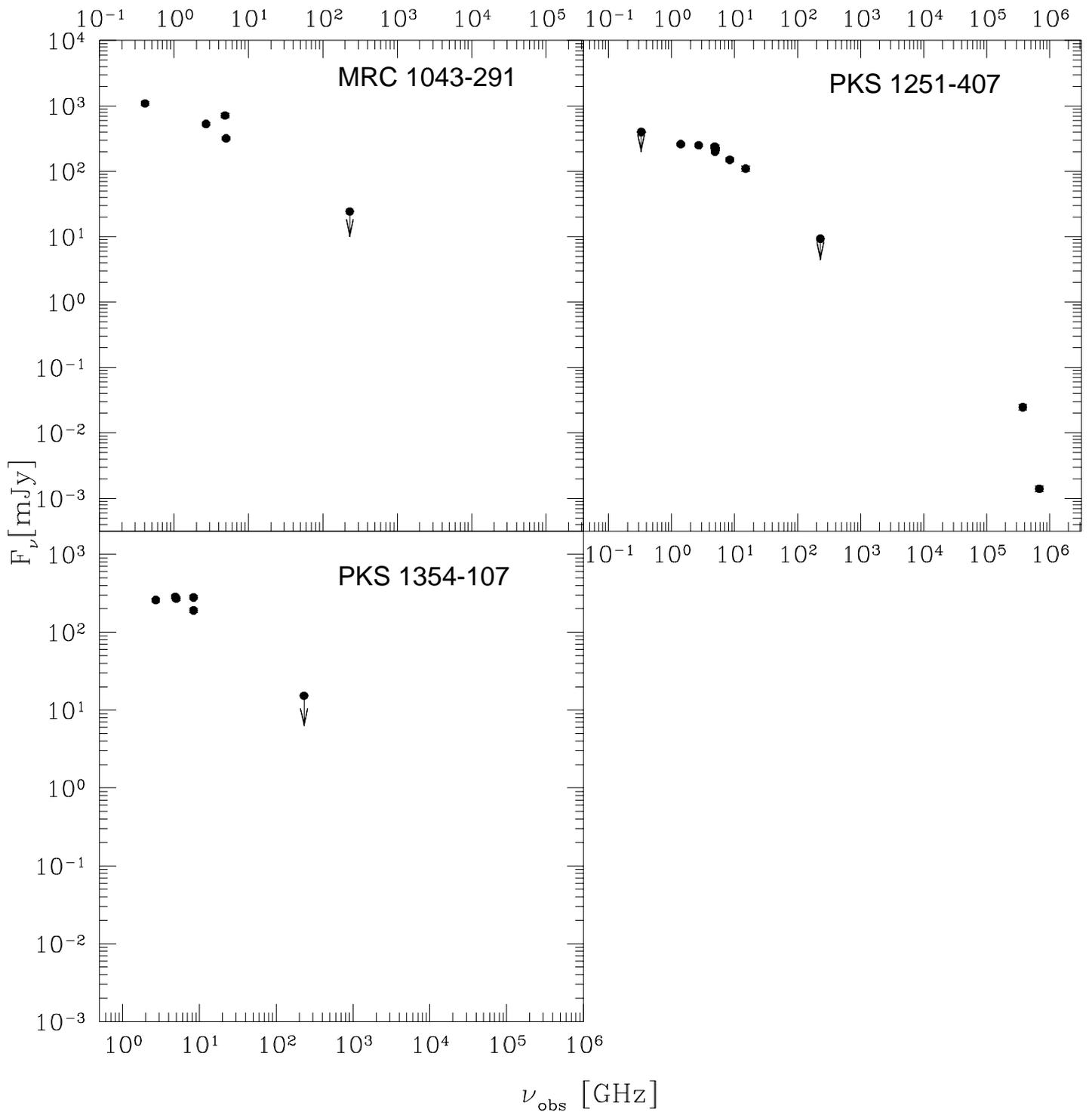
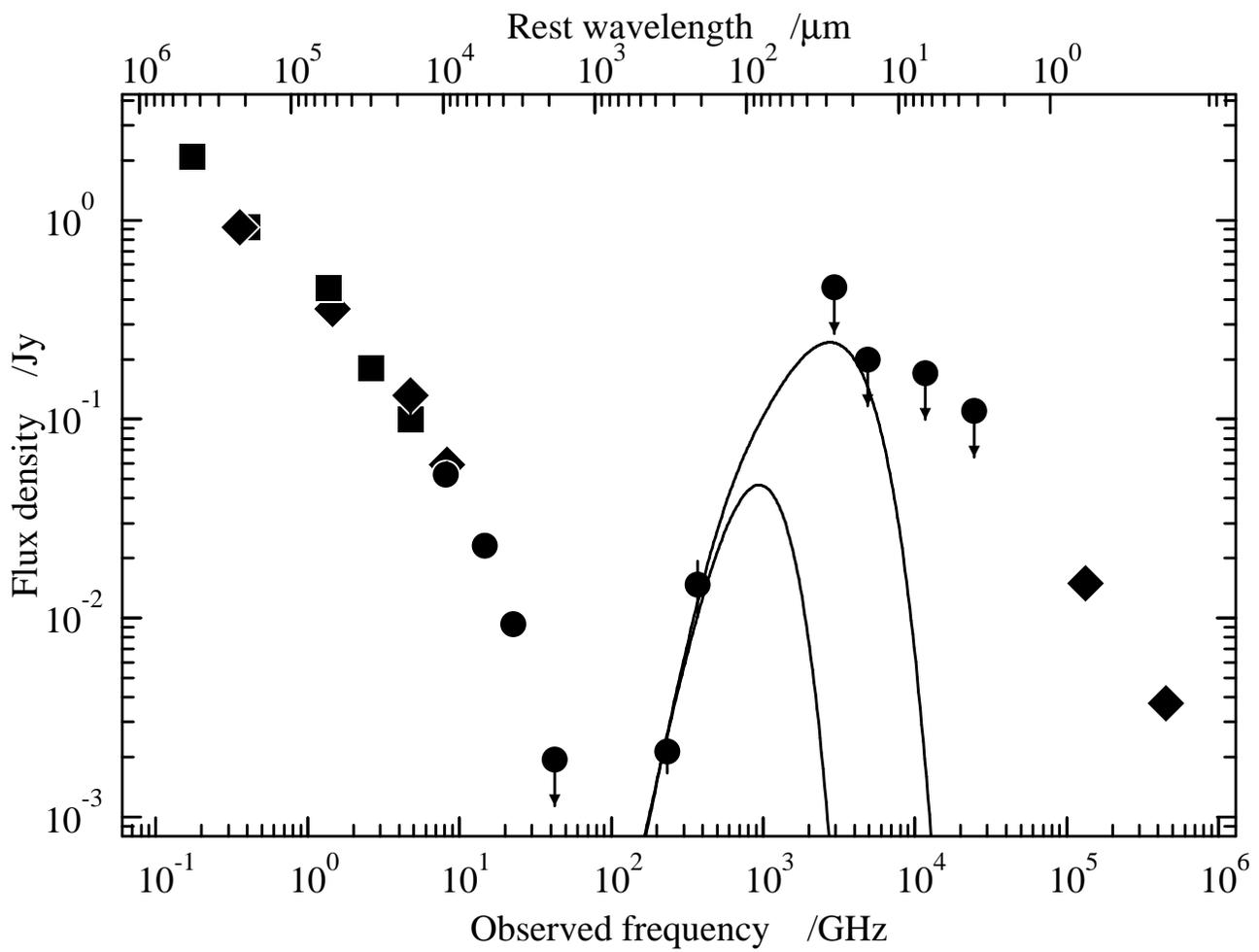
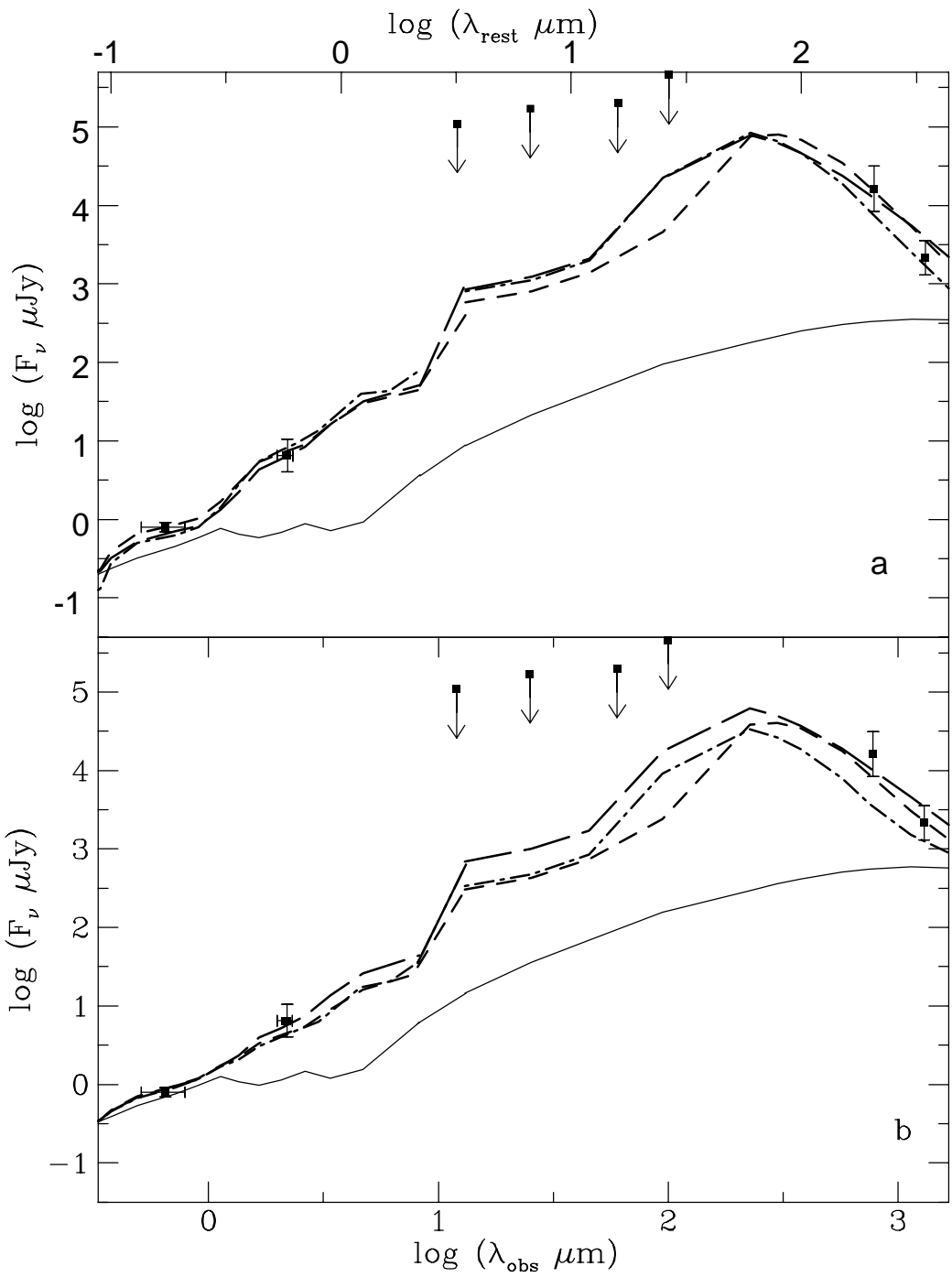
\end{document}